\documentstyle[twocolumn,seceq,epsbox]{jpsj}
\newcommand{\be}{\begin{equation}}
\newcommand{\ee}{\end{equation}}
\newcommand{\bd}{\begin{displaymath}}
\newcommand{\ed}{\end{displaymath}}
\newcommand{\ba}{\begin{eqnarray}}
\newcommand{\ea}{\end{eqnarray}}

\def \theequation{\thesection.\arabic{equation}}
\def\gsim{\lower3pt\hbox{$\stackrel{>}{\scriptstyle \sim}$}} 
\def\lsim{\lower3pt\hbox{$\stackrel{<}{\scriptstyle \sim}$}} 

\def\nnu{\mib{\nu}}
\def\mmu{\mib{\mu}}
\def\u{\mib{u}}
\def\f{\mib{f}}
\def\w{\mib{w}}
\def\x{\mib{x}}
\def\X{\mib{X}}
\def\V{\mib{V}}
\def\r{\mib{r}}
\def\R{\mib{R}}
\def\k{\mib{k}}
\def\q{\mib{q}}

\def\e{{\rm e}}
\def\d{{\rm d}}
\def\St{\tilde{S}}
\def\no{\nonumber}

\def\av#1{{\left\langle#1\right\rangle}}
\def\pa#1#2{\frac{\partial #1}{\partial #2}}
\def\paa#1{\frac{\partial}{\partial #1}}
\def\ave{\overline{\varepsilon}}

\newcounter{fgr}
\begin{document}
\def\runtitle{Structure Functions in Decaying and Forced Turbulence}
\def\runauthor{Daigen {\sc Fukayama}, Toshihiro {\sc Oyamada}, Tohru {\sc
Nakano},
Toshiyuki {\sc Gotoh}, and Kiyoshi {\sc Yamamoto}}
\title
{
Longitudinal Structure Functions in Decaying and Forced Turbulence}

\author
{
Daigen {\sc Fukayama}$^{1}$
\footnote{E-mail address: daigen@phys.chuo-u.ac.jp},
Toshihiro {\sc Oyamada}$^{1}$,
Tohru {\sc Nakano}$^{1}$,  \\
Toshiyuki {\sc Gotoh}$^{2}$ and Kiyoshi {\sc Yamamoto}$^{3}$
}

\inst
{
$^1$ Department of Physics, Chuo University, Tokyo 112-8551\\
$^2$ Department of Systems Engineering, Nagoya Institute of Technology,
Nagoya 466-8555\\
$^3$ National Aerospace Laboratory, Chofu, Tokyo 182-8522
}

\recdate
{
}

\abst
{
In order to reliably compute the longitudinal structure functions in decaying
and forced turbulence, local isotropy is examined with the aid of the
isotropic expression of the incompressible conditions for the second and third
order structure functions. Furthermore, the Karman-Howarth-Kolmogorov relation is investigated to examine the effects of external forcing and
temporally decreasing of the second order structure function.
On the basis of these investigations, the scaling 
range and exponents $\zeta_n$ of the longitudinal structure 
functions are determined for decaying and forced turbulence with the aid of 
the extended-self-similarity (ESS) method. 
We find that $\zeta_n$'s are smaller, for $n \geq 4$, in decaying turbulence
than in forced turbulence. The reasons for this discrepancy are discussed.
Analysis of the local slopes of the structure functions is used to
justify the ESS method.}

\kword
{Karman-Howarth-Kolmogorov equation, isotropy, external forcing,
non-stationarity, structure functions, longitudinal velocity increment}

\sloppy
\maketitle


\section{Introduction}

Study of structure function is useful in the analysis of the scaling property
of fully-developed turbulence.~\cite{rf:my}  In particular, the structure
function of the longitudinal velocity increment clearly shows how the scaling
deviates from K41 scaling~\cite{rf:k} as the order of the structure
function increases.~\cite{rf:sl} The structure function of a transverse
velocity increment also exhibits deviation from K41 scaling,
but the values of the scaling exponents appear to differ from
those of the longitudinal variety, reflecting the influence of vortical
structures in turbulence.~\cite{rf:camussi1,rf:noullez,rf:camussi2,rf:boratav,rf:chen,rf:dhruva}

All measurements in direct numerical simulations (DNS's) and
experiments include certain limitations or restrictions. Examples of these
restrictions are the limitation of the Reynolds number to finite values, the size
of the samples, and the degree of deviation from isotropy, homogeneity, and
stationarity that are allowed to the system. Therefore the results obtained from such data
must be studied carefully and thoughtfully.
To be more specific, consider the DNS of forced homogeneous isotropic
turbulence with a finite Reynolds number, in which the random forces are
assumed to be statistically homogeneous, isotropic and Gaussian
with a given spectrum. In order to realize a high Reynolds turbulence number,
the force spectrum is assumed to have compact support at the very low
wavenumber range, $1\le k\le 2\sim 3$. This restriction introduces a problem
of convergence to large scale values with respect to isotropy, homogeneity
and stationarity.
Since the small number of Fourier modes in the forced band have 
a large amount of kinetic energy, and their characteristic times are long, 
fluctuation of these
modes significantly affects the statistics of large-scale turbulence,
which possibly pertain to the inertial region.  The difficulty lies
in determining for how long statistical quantities must be averaged to assure
statistical homogeneity, isotropy and stationarity of the system,
and in discovering the largest scale not affected by the
external forces.

In simulation of decaying turbulence, similar problems arise, because
the fluctuations of the Fourier modes in the lower band are large at $t=0$.
Hence the issues to be addressed are the effect of non-stationarity on
the intensity of fluctuations, the anisotropic effect due to large
eddies, and the size of the largest scale that is not influenced by a decrease in the intensity of fluctuations.

In laboratory tests, such as experiments using the 
turbulence in a cylinder with disks rotating in opposite directions,~\cite{rf:mtw} 
we find a situation similar to the case of forced DNS. The
geometry of the experimental apparatus imposes certain limitations on the isotropy
and homogeneity of the system,
although the sample size is relatively larger than in the DNS case and the Reynolds number
is relatively lower than that in the case of atmospheric flow.  Also, the total
record length is usually much longer than the large-eddy turnover-time.
In an atmospheric flow, on the other hand, the Reynolds number is
usually very large, but the characteristic time of
the macroscale shear quite often becomes comparable to or longer than the
record length. Thus, the homogeneity and isotropy conditions are not well satisfied,
and the problem of statistical convergence at large scales
arises.~\cite{rf:dhruva}

Most studies of the scaling of structure functions in the DNS
have been made using a very limited number of samples,
and the results may therefore suffer from the effects discussed above.
Hence, there are several factors to be checked before drawing definite conclusions
from simulations about scaling; they are: (1) the degree of isotropy of turbulence
in the scaling region, (2) the effect of external forcing, and (3) the effect
of decreasing intensity of fluctuations.  Forced simulation cannot
avoid the first two factors, while the first and third factors are crucial in
decaying simulation.  Thus, we need to investigate the effects
of limiting the Reynolds number to finite values, and then determine the 
resulting longitudinal structure function.

In the present paper we carefully examine the above three factors in decaying
and forced simulated turbulence with various Reynolds numbers.  In particular,
flows of Reynolds number $R_{\lambda}=70$ are carefully studied.  Twelve
decaying
turbulences with the same initial energy spectrum (but different realizations) are
simulated, and the forced turbulence is executed for a long period of time,
providing an average
over 126 samples, during approximately 50 eddy turnover times.
The degree of local isotropy is studied through the incompressible-condition
restriction on the second and third order structure functions.  How external
forces penetrate into the upper inertial region in forced turbulence and how the
non-stationarity of the fluctuations deteriorates the upper inertial region in
decaying turbulence are investigated through the Karman-Howarth-Kolmogorov relation
(referred to as the KHK relation hereafter).~\cite{rf:kh} The KHK relation is an exact
expression connecting the third-order structure function to the second order structure function.
On the basis of these studies, we estimate the scaling exponents of the longitudinal
velocity structure functions in forced and decaying turbulence by employing
the extended-self-similarity (ESS) method.~\cite{rf:benzi1,rf:benzi2}  We find
that the scaling exponents of the forced turbulence agree with other currently
reported values,~\cite{rf:sl} but that the scaling exponents of decaying
turbulence are smaller for higher orders than the corresponding exponents of 
steady turbulence.  A reason for this behavior is suggested using the
equation for the structure function of arbitrary order, which is a generalization
of the third-order structure function.

Our paper is organized as follows.  In section 2, the relevant data from
several simulations of decaying and forced turbulence are summarized.
Section 3 is devoted to discussion of the isotropy of turbulence,
while the KHK relation is investigated in section 4.  In section 5 the
longitudinal structure function is obtained in the scale range that is appropriate
for the KHK relation and the conditions of isotropy to hold.  The scaling
exponents of longitudinal structure functions up to the eighth order are
derived using the ESS method.  Evidence that the exponents for
decaying turbulence are smaller than those for forced turbulence is also presented.
In section 6, the effect of the finite Reynolds number on the structure of the local 
exponents is discussed, and a justification for the ESS method is given. 
In the appendices, the equation for the structure function of arbitrary order is
derived and, in particular, the second-order equation is shown to reduce to the
KHK relation. Additionally, the contribution of the pressure gradient term in the equation
for the structure function is shown to be short ranged.~\cite{rf:lp}

\section{Numerical Procedure}

Three kinds of decaying runs were carried out for the Gaussian, random,
initial-velocity field, using the same energy spectrum $E(k,0)=c(k/k_0)^4 \exp[-2(k/k_0)^2]$($k_0=3$ for Run 1 and $k_0=1$ for Run 2 and Run 3),
but with different realizations. 
The number of mesh points used were $256^3$ for Runs 1 and 2, and $512^3$
for Run 3.
Run 1 consisted of 12 different sub-runs with the same Reynolds number.
The numerical algorithm used for the calculations is the pseudo-spectral method
in space, and the fourth-order Runge-Kutta-Gill method in time. Aliasing errors
were eliminated through application of the shift method. The computations were
carried out on a parallel vector-machine at NAL's Numerical
Wind Tunnel and at RIKEN's Advanced Computing Center.
The various data were taken at the time where the energy dissipation
had reached the maximum value. The relevant data are shown in Table \ref{table:1}.
The spatial resolution of the DNS's may be estimated by the value
of the product $k_{\mbox{max}}\eta$, where $\eta$ is the Kolmogorov length.
For all runs, the values obtained are larger than unity, implying that sufficient 
resolution was obtained for even the smallest scale.

The numerical method used for forced turbulence is essentially the same as
that used for decaying turbulence. The initial conditions applied
to the forced-turbulence experiment were also the same as those used in the
decaying-turbulence test. The random force is statistically homogeneous,
isotropic and Gaussian white, and its spectrum support is limited to the band
$ 2 \lsim k \lsim 3$. Also, the spectrum form is constant in the band.
The data are shown in Table \ref{table:1}.
After the transient period had passed, usually after a few turnover times,
steady states were attained and the time average was computed 
over some multiple of the turnover time. For example, Run 4 was averaged over 126 samples
during 50 eddy turnover times, and Run 5 was averaged over 45 samples during 9 eddy
turnover times. The computations were performed at RIKEN's Advanced Computing Center.

\begin{table}[t]
\renewcommand{\arraystretch}{1.0}
\begin{center}
\begin{tabular}{@{\hspace{\tabcolsep}\extracolsep{\fill}}cccccc} \hline
$ \enskip$    & Run 1 & Run 2& Run 3 & Run 4 & Run 5 \\ \hline
condition & decay & decay & decay & forced & forced \\
mesh points    & $256^3$ & $256^3$ & $512^3$ & $256^3$ & $512^3$  \\
$R_\lambda$  & $70$   & $90$  & $120$  &  $70$ & $125$  \\
$k_{\mbox{max}}\eta$ & $1.11$  & $1.26$  & $1.42$  & $2.32$ & $2.03$ \\
$ N $  & 12 & 1 & 1 & 126 & 45   \\ \hline
\end{tabular}
\end{center}
\vskip -0.2cm
\caption{Various parameters in decaying and forced simulations. $N$ is the
number of samples over which the average is taken.}
\label{table:1}
\end{table}
\vskip -0.5cm

\section{Isotropy}

There are many ways to check the isotropy of turbulence.
One is to examine the relation $\av{u^2}=\av{v^2}=\av{w^2}$, which reflects the
isotropy of the largest scale, or the ^^ global" isotropy.  In order to check
the degree of local isotropy of eddies of scale $r$, which is necessary to compute
the scaling exponents of the structure functions, we study the isotropic nature of
the second and third order structure functions.  When the incompressible
condition is applied to the second and third order structure functions, the
following relations result for the isotropic case:~\cite{rf:my}
\ba
   D_{TT}(r) &=& D_{LL}(r)+\frac{r}{2} \frac{\d D_{LL}(r)}{\d r}, \label{2-1} \\
   D_{LTT}(r) &=& \frac{1}{6} \frac{\d}{\d r}rD_{LLL}(r), \label{2-2}
\ea
where $D_{LL} \equiv \av{(\delta u_l)^2}, \, D_{TT} \equiv \av{(\delta u_t)^2},
\, D_{LLL} \equiv \av{(\delta u_l)^3}, \, D_{LTT} \equiv \av{\delta u_l (\delta
u_t)^2}$, $\delta u_l$ is the longitudinal velocity increment over a distance $r$, and
$\delta u_t$ is the transversal velocity increment.  Instead of the differential
forms shown above, in the following derivations we
employ the integral forms
\ba
  & & \int _0^r \left( D_{TT}(r') -\frac{1}{2} D_{LL}(r') \right) \d r'
               =\frac{1}{2}r D_{LL}(r) , \label{2-3} \\
  & & \int_0^r D_{LTT}(r')\d r'= \frac{1}{6}rD_{LLL}(r), \label{2-4}
\ea
because numerical computation is less noisy for the integrals than
for the derivatives.

\subsection{Decaying turbulence}

Figures 1(a) and 1(b) show comparisons with varying $r$ for Run 1, averaged
over 12 samples of different realizations. This shows that the above relations
are satisfied. The relation (\ref{2-4}) for the third-order structure
function is less accurate than that for the second-order, but
it still holds up to $r_* \equiv r/\eta=160$.  (Note that from now on,
the scale $r$ will be expressed in units of the Kolmogorov length $\eta$.)
The local isotropy of the structure functions is improved with an increase
in the number of samples. To witness this, consider the isotropic relation
for Run 2, consisting of a single sample. In this case, the third-order relation
was found to be slightly violated beyond $r_* =30$ (figures not shown).

Later we will determine the scaling indices of the longitudinal structure
functions in the inertial range using the ESS method. We will employ the scale range
$16.0 \leq r_* \leq 29.4$ for Run 1. Because of this choice of scale range, the structure functions in the inertial region are not affected by any lack of
accuracy in the degree of the isotropic condition of the turbulence.  This is the case for Runs 2 and 3.

Although it is not as easy to check the isotropy of the higher-order structure functions
as it is to check the second and third-order functions,
we expect the statistical convergence for the isotropy of
the structure functions to be gradually lost as the order of the function increases.
This is because the higher-order structure functions are dominated by rare, strong
events that are oriented in specific directions. However, this does not necessarily
mean that the degree of anisotropy of the fourth-order structure function is
larger than that of the third-order structure function. The statistical convergence of even-ordered
structure functions is faster than that of odd-ordered structure functions.

In decaying turbulence, the state of isotropy is satisfactorily established.
Although a sufficient number of samples is desirable for 
computation of the structure functions, the data of a single snap-shot (as in Run 2 or 3) are enough to determine the structure functions with respect
to the degree of isotropy. This, however, is not the case with forced turbulence, as is discussed below.

\begin{figure}
\epsfile{file=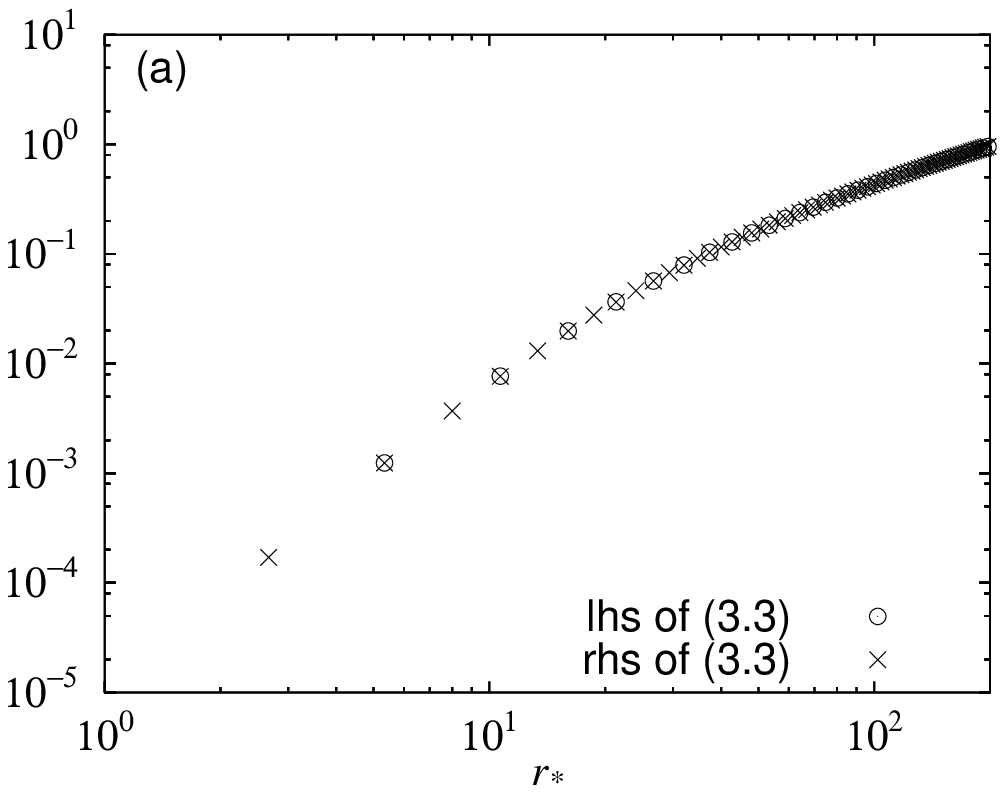,width=8.0cm,height=6.0cm}

\epsfile{file=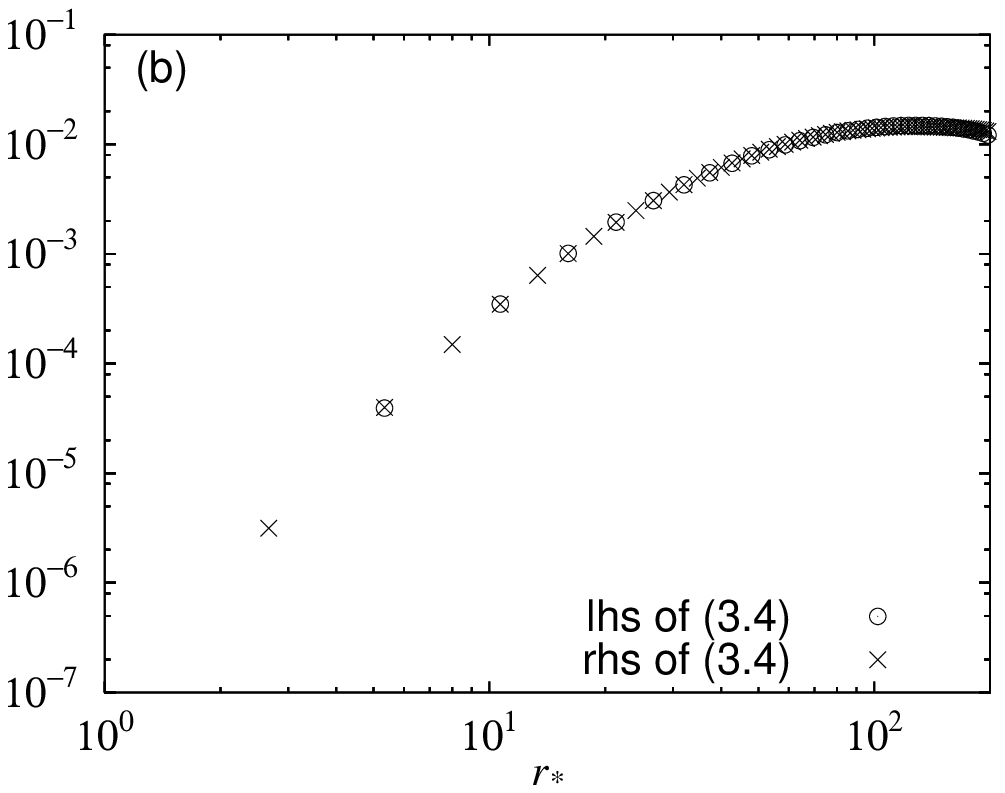,width=8.0cm,height=6.0cm}
\vspace{-8mm}
\caption{
The isotropic conditions for Run 1, consisting of
12 identical samples with different realizations.  (a) The second-order
structure function; (b) the third-order structure function.

}
\end{figure}

\subsection{Forced turbulence}

It is important to note that the conditions (\ref{2-3}) and (\ref{2-4})
are not satisfied for forced turbulence as well as they are for decaying
turbulence, particularly when the data of only one snap shot are processed.
To see this explicitly, we show in Figs. 2(a) and 2(b) the curves of eqs.
(\ref{2-3}) and (\ref{2-4}) for Run 4 at a certain instant after the
turbulence reaches the steady state.  Although the second-order structure
function satisfies the isotropic condition, the same condition applied to the
third-order structure function is violated at all scales. This indicates that
an average over a prolonged period is needed to establish 
the degree of isotropy for forced turbulence. 
On the other hand, Figs. 3(a) and 3(b)
show curves for Run 4 averaged over 126 samples, with the averaging-time being
approximately 50 eddy turn-over times. The isotropy condition for the second-order
structure function is thus ensured quite satisfactorily. The isotropy condition
for the third-order structure function, however, is only satisfied up to
$r_*=50$.  We expect that the condition for statistical isotropy becomes better
satisfied as the number of samples increases, and as the averaging-time increases.
This would suggest that steady turbulence with isotropic forcing needs to be
studied over a sufficiently prolonged averaging time.

\begin{figure}
\epsfile{file=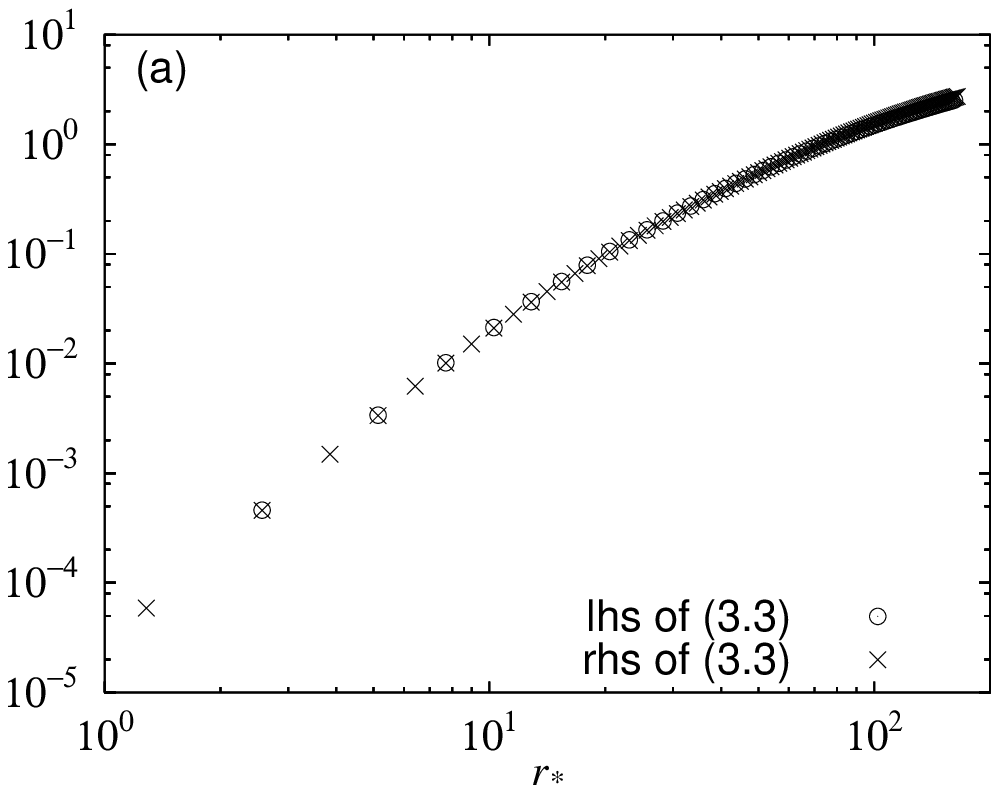,width=8.0cm,height=6.0cm}

\epsfile{file=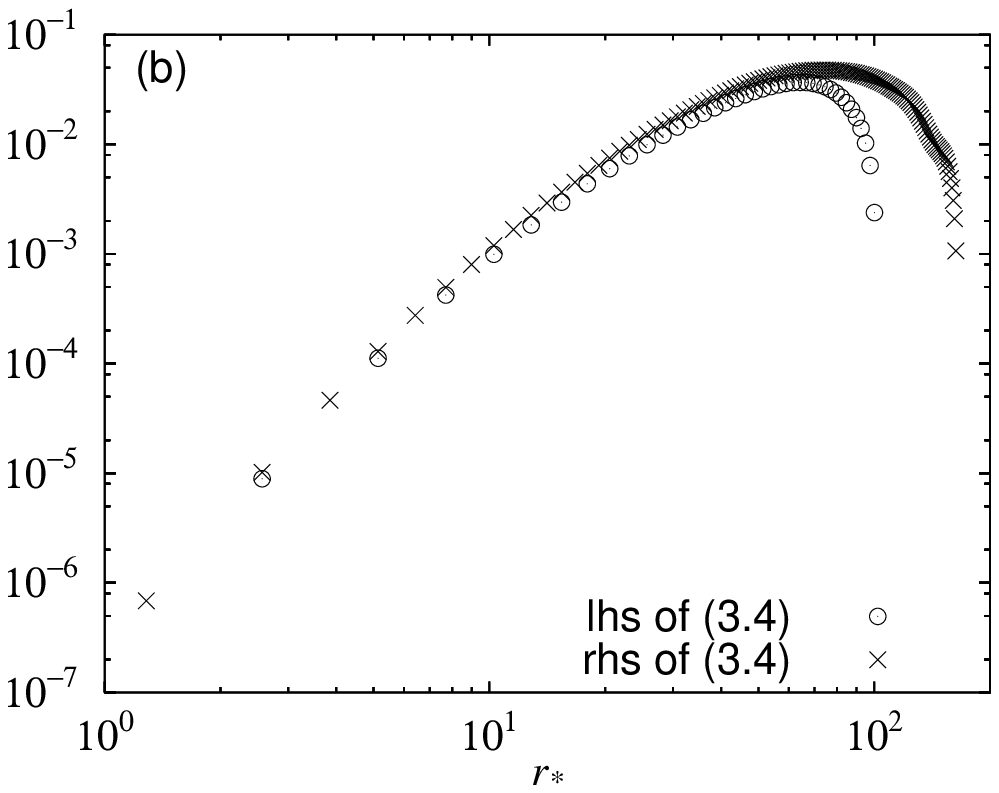,width=8.0cm,height=6.0cm}
\vspace{-8mm}
\caption{
The isotropic conditions at one snap-shot for Run 4, when the energy
dissipation reaches a constant level. (a) The second-order structure
function; (b) the third-order structure function.

}
\end{figure}

\begin{figure}
\epsfile{file=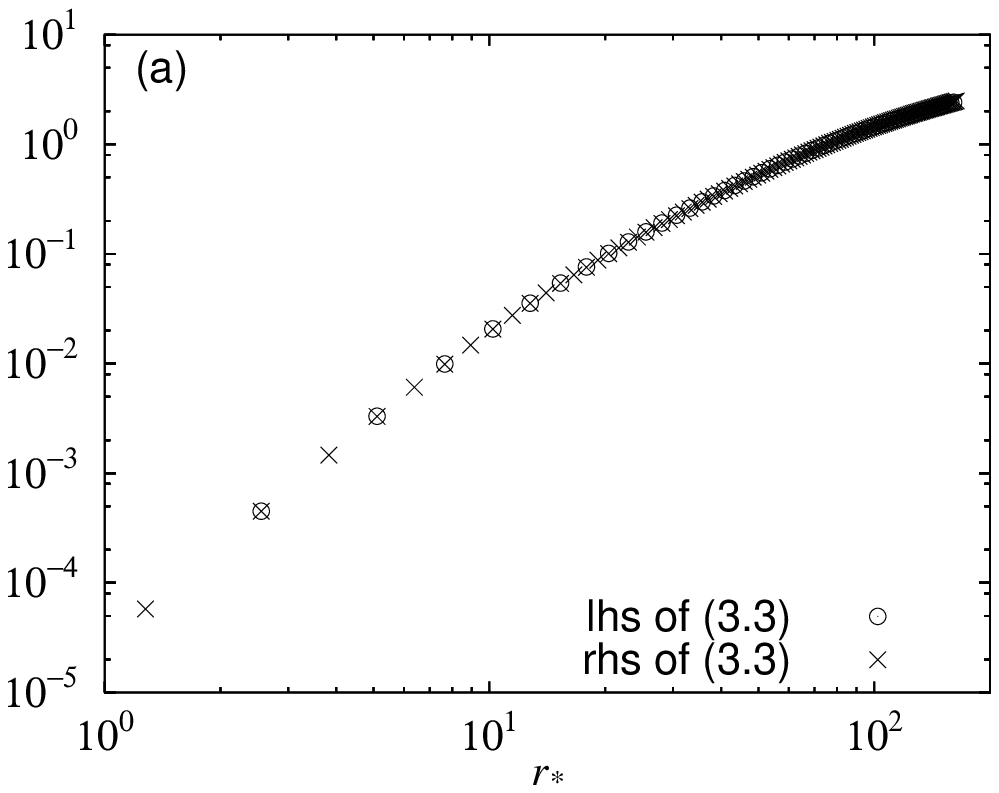,width=8.0cm,height=6.0cm}

\epsfile{file=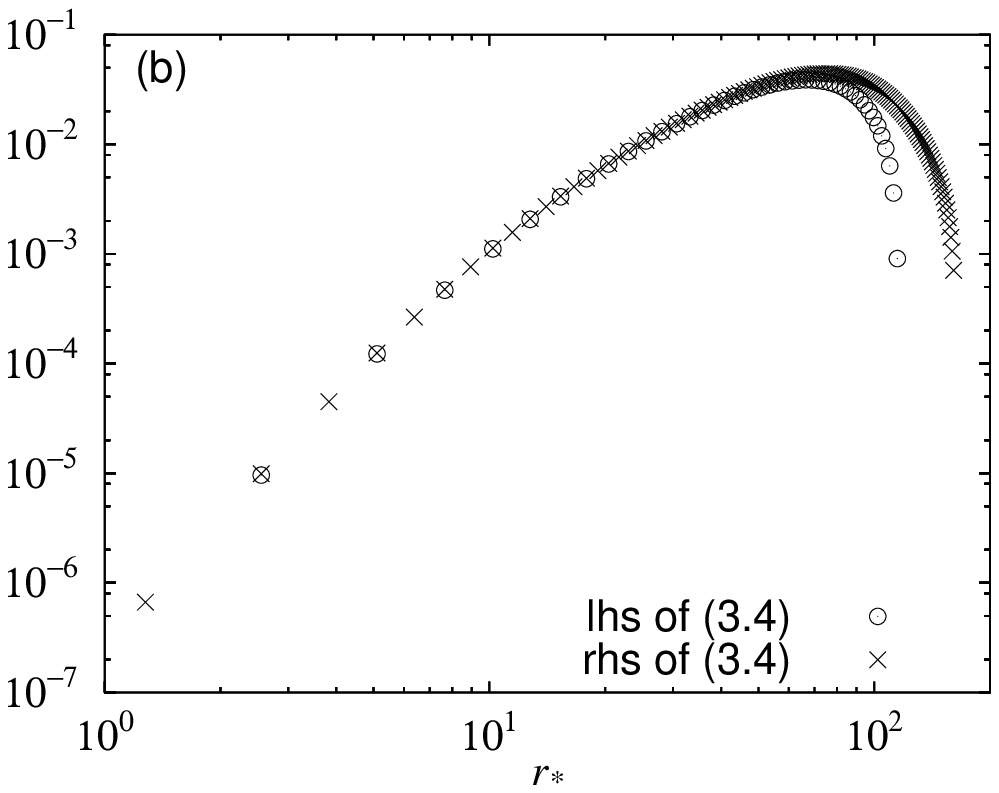,width=8.0cm,height=6.0cm}
\vspace{-8mm}
\caption{
The isotropic conditions for Run 4 averaged over 126 samples during
approximately 50 eddy-turnover times. (a) The second-order structure function;
(b) the third-order structure function.
}
\end{figure}

\section{The Karman-Howarth-Kolmogorov Equation}

In this section we consider, through the KHK relation~\cite{rf:kh}, how 
external forcing (in forced turbulence) and non-stationarity of the intensity
of fluctuations (in decaying turbulence) affect the transfer of energy in
scale-space. The effects of the external forcing and non-stationarity on a weighted transfer of energy,
which is the average of the product of the energy-dissipation rate
and the multiples of the velocity-increment, associated with the higher-order
structure functions are discussed later.

The KHK relation is an exact equation, which is very rare in turbulence theory.
The equation for the second-order structure function employing the isotropic
condition is written:
\be
  D_{LLL}(r)=-\frac{4}{5}\ave r + 6\nu \pa{D_{LL}}{r} + G(r)+F(r), \label{4-1}
\ee
where
\be
    G(r)=-\frac{3}{r^4} \int^r_{0}\pa{D_{LL}}{t}r'^4 \d r', \label{4-20}
\ee
and
\be
    F(r)=\frac{2}{35} \varepsilon_{in}(k_e r)^2 r, \label{4-21}
\ee
as is derived in eq. (\ref{a-9}) of Appendix A.  Note that equation (\ref{4-1})
is valid for decaying turbulence as well as for the forced variety.  Here $\ave$ is the
average dissipation rate, while $\varepsilon_{in}$ is the energy input rate
due to external random forces, which are distributed about $k \sim k_e$.
Note that $\nu$ represents the kinematical viscosity. The second term on the
right-hand side (RHS) of eq. (\ref{4-1}) expresses the effect of viscous damping,
while the third term on the RHS relates the effect of the non-stationarity (decreasing
in time) of the second-order structure function, which is significant for decaying
turbulence. The fourth term on the RHS expresses the effect of the external forces.
Note that only in steady, forced turbulence does $\ave=\varepsilon_{in}$.
It is of some interest to note that the first term on the RHS of eq. (\ref{4-1})
is negative, while the viscous term, the external-forcing term
and the non-stationarity term are positive.

In the inertial region of stationary-turbulence with infinite Reynolds number,
one can neglect the second, third and fourth terms on the RHS of
the above equation, so that one gets Kolmogorov's well-known 4/5
law~\cite{rf:k2}:
\be
  D_{LLL}(r)=-\frac{4}{5}\ave r . \label{4-1a}
\ee
The inertial region in real flows with finite Reynolds numbers is regarded as
having the relation (\ref{4-1a}) hold approximately, and hence the universal scaling
property of the structure functions in the inertial region must be examined
only under these conditions.  In actual flows, however, the Reynolds number is not
large enough to guarantee the wide spatial range in which eq. (\ref{4-1a}) will hold.

For flows of finite Reynolds numbers, therefore, the important issue is how the
transfer term $-(4/5)\ave r$ is affected by (1) the viscous damping term,
(2) the non-stationarity term in decaying turbulence and (3) the external forcing term
in forced turbulence. Case (1) is significant in the lower inertial region, while (2)
and (3) are pertinent in the upper inertial region.  We investigate each term in the
KHK equation for both the decaying- and
forced-turbulence scenarios.  The failure of eq. (\ref{4-1}) is caused by the
breakdown of the isotropic condition, and by the insufficient number of samples 
averaged.

Now, concerning the numerical computation, $\ave$ was calculated from the square of the
spatial derivative of the velocity field, while the forcing term
$F(r)$ in eq. (\ref{4-21}) was computed with the aid of the spectrum of
the random forces, not from an external energy input at each instant.
More precisely, we did not use the approximation $F(r)$ for the
forcing term, but the exact expression (\ref{a-10}) in Appendix A; we did this
because, in this case, $k_e r$ is not limited to $k_e r \ll 1$. The other terms
in the expression are all evaluated directly through processing the data.

\subsection{Decaying turbulence}
The Karman-Howarth-Kolmogorov relation now takes the form
\be
  D_{LLL}(r)=-\frac{4}{5}\ave r
        +6\nu \frac{\partial D_{LL}(r)}{\partial r}+G(r). \label{4-2}
\ee
In Fig. 4, we depict $-D_{LLL}(r)/\ave$ and the combined sums of the
terms on the right-hand side of eq. (\ref{4-2}) divided by $\ave$, against the scale
$r_*$ for Run 1 (averaged over 12 samples.)  The relation (\ref{4-2}) is well
satisfied for all values of $r_*$.

From Fig. 4 it can be seen that the peak of $-D_{LLL}(r)/(\ave r)$ is 0.53 at $r_*=26.7$,
and the linear region of $D_{LLL}(r)$ is extremely narrow around $r_* \sim 23$.
If the viscous effect is included, however, $-D_{LLL}(r)+6\nu \partial D_{LL}/\partial r$
is in near-agreement with $(4/5) \ave r$ up to $r_* =10$.  If the non-stationarity
of $D_{LL}$ is taken into account, agreement is guaranteed for any scale.
The effect of the non-stationarity of $D_{LL}$ becomes more significant
as the separation scale $r$ increases, indicating that the amplitudes of
larger-scale components decay faster than those of smaller-scale components. This is because
the former do not have any energy supplied to them in the decaying turbulence.
The non-stationarity term, which has been included in Fig. 4, is observed to
decrease as $r^{c}$ for $ 5 \leq  r_* \leq 30$, and where $c$ is almost 3.
This constant slope is surprising, because the scaling range of the
non-stationarity term extends from the inertial to the dissipation region.

We will now discuss the form of the KHK relation in connection with the structure
function, with the details of the computations shown in the following section.
As seen from Fig. 4, corresponding to Run 1, it is difficult to find the region
where $D_{LLL}(r)=-(4/5) \ave r$. This implies that computation of
the exponents in the inertial region by the direct-fitting method is not possible.
We therefore employ the ESS method~\cite{rf:benzi1,rf:benzi2}
to calculate the scaling exponents of the structure functions.  The ESS method is known to extend
the scaling region down to the upper dissipation range~\cite{rf:sreen},
because the third-order structure function used in the abscissa contains
the effect of the viscous-damping term.  In the following section,
on the other hand, we will show that the ESS method yields reasonable values
beyond $r_*=20$ up to $r_*=30$, where, as Fig. 4 shows, the non-stationarity
effect cannot be neglected. Thus the ESS method extends the scaling region
upward in the scale-space in decaying turbulence.

The KHK equation has been experimentally investigated in grid-turbulence
by Danaila {\it et al}.~\cite{rf:danaila}, who took into account the
non-stationarity term by approximating it using spatial differentiation
under the Taylor hypothesis.  They claimed that the KHK equation is correct
up to $r_* \sim 200$; however a slight discrepancy is seen around $r_* \sim 20$
(Fig. 3 in ref. 17). Such a deviation negatively influences the determination of
the scaling exponents of the higher-order structure functions.
The non-stationarity term takes the form $r^c$ with $1<c<2$, in contrast
to our assumed value of $c=3$.  The reason for this discrepancy is not currently
known.

\begin{figure}
\epsfile{file=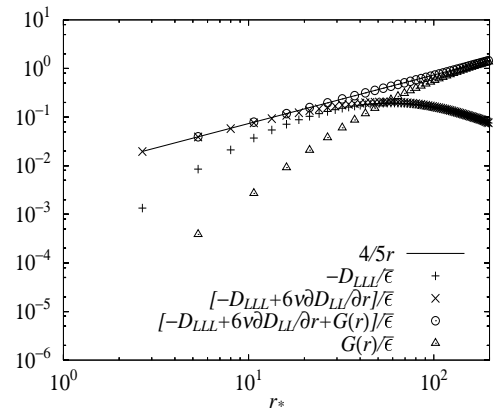,width=8.0cm,height=6.0cm}
\vspace{-8mm}
\caption{
The KHK relation for decaying turbulence for Run 1 averaged over 12
samples.
}
\end{figure}

\subsection{Forced turbulence}

In this case the KHK equation takes the form
\be
  D_{LLL}(r)=-\frac{4}{5}\ave r + 6\nu \pa{D_{LL}(r)}{r} +F(r), \label{4-3}
\ee
where the last term is numerically computed with the help of the exact
expression (\ref{a-10}) in Appendix A.  It is interesting to note
that the relation (\ref{4-3}) does not hold when the data of only a single
instantaneous snap-shot are processed. This situation is analogous to having
the isotropic condition fail in Fig. 2. The reason for the similarity is that
both are linked to a random feature in the external input. The last term in
eq. (\ref{4-3}) is calculated using an ensemble-average value from the
spectrum of $\av{|f_i(\k)|^2}$, which is equivalent to a time average over a
sufficiently long period. Let us now estimate how the input term varies
from time to time.  The number of input modes is $N=160$ for Runs 4 and 5, so that
a simple estimate of the relative amplitude of the input fluctuation is of the order
$1/\sqrt{N} \sim 0.1$.  This implies that the external-forcing term at 
any given instant may have a relative error on the order of 10\%.  In order to
obtain reliability of the forcing term to within a 1\% error, we need $10^4$
random forces, which corresponds to roughly one eddy-turnover time in our present
simulation. To obtain the statistically meaningful result of eq. (\ref{4-3}) 
we require a much longer averaging period than over only one eddy-turnover time.

Figure 5 displays the KHK curve for Run 4 with its average over 126 samples,
in a time interval of roughly 50 eddy-turnover times.  It is not surprising that the
KHK relation holds here for almost any separation, since the relation is exact
under the isotropic condition. Let us now inspect the contribution of each term
in the KHK equation to its total value. According to Fig. 5, the peak of
$-D_{LLL}(r)/(\ave r)$ is 0.58 at $r_*=26.8$, and the flat region of $D_{LLL}(r)/r$
is extremely narrow around $r_* \sim 26$. This is very similar to the situation
encountered in the decaying turbulence studied in Run 1. If the viscous effect is also
taken into account however, the agreement of $-D_{LLL}(r)+6\nu \partial D_{LL}/\partial r$
with $(4/5)\ave r$ is nearly correct up to $r_* =20$. If the effect due to
external forces is included, which begins to appear at $r_* \sim 20$, the agreement
between the two relations is perfect.

Also of interest are the results of Run 5, averaged over 45 samples, during
nine eddy-turnover times. Here the KHK relation is not obeyed as well as in Run 4.
Our computations confirm that agreement improves as the sample number
increases, which explains why the agreement is better for Run 4. Note however
that the improvement in agreement is small with modest increases in sample number.

The KHK relation for a real flow was recently studied~\cite{rf:mtw} for low
temperature helium gas flows, with $R_{\lambda}$ ranging from 120 to 1200, using
an apparatus in which the turbulence is forced by two counter-rotating disks in a
cylindrical container.  The effects of large-scale motions were included in
eq. (\ref{4-3}) as the external-forcing term.  The exact nature of eq. (\ref{4-3})
was proved in this experiment, except for large scales, where non-isotropic
forcing has a dominant effect.

\begin{figure}
\epsfile{file=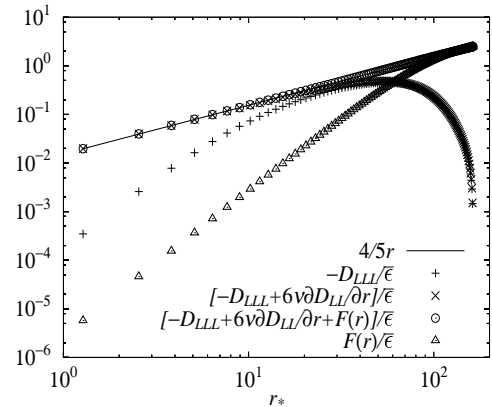,width=8.0cm,height=6.0cm}
\vspace{-8mm}
\caption{
The KHK relation for Run 4 averaged over 126 samples.
}
\end{figure}

\section{Longitudinal Structure Functions}

Let us begin with an estimate of the maximum degree of the order of structure
functions that we desire to measure accurately. We plot the probability-density
function $P(u_r)$ multiplied by $u_r^n$ against $u_r$, where $u_r$ is the
longitudinal-velocity increment over the distance $r$. As $n$ increases,
the peak of the bell shape of $u_r^nP(u_r)$ approaches the larger value of $|u_r|$.
Beyond a certain value of $n$, the peak cannot be distinguished and rare
strong events heavily influence the computation of $\av{u_r^n}$.
Figure 6 is an example of such a plot, showing $u_r^8 P(u_r)$ at $r_*=20.8$
(Run 3). The abscissa in the plot is $u_r/\sqrt{S_2(r)}$, while the ordinate
is pre-divided by $S_2(r)^4$. The curves decay quickly for large amplitudes. 
For the plots of $u_r^nP(u_r)$ with $n\ge 9$, we observed that the 
decay of the curves is not sufficiently fast to ensure that moments 
higher than the ninth order are computed accurately (figure not shown). 
Thus we conclude from these simulations that $n=8$ is the maximum order allowed
for the condition of a sufficient-number-of-events to be met in the statistical sense.

In this paper we will concentrate our efforts only on the study of the 
longitudinal-velocity structure function
\be
    S_n(r)=\av{u_r^n}.  \label{5-1}
\ee
Before going on to determine the exponents of $S_n(r)$, we will examine
the structure functions of odd orders.  For this purpose, let us recall the
probability-density function (PDF) of $u_r$: $P(u_r)$. The peak of the PDF
is located at the small positive value of $u_r$, while its tail is more
populated on the negative side than on the
positive side, in accordance with $\av{u_r}=\int u_r P(u_r)\d u_r=0$. For
$\av{u_r^3}$, on the other hand, the contribution from the negative values
of $u_r$ is larger than that from the positive values, implying a negative
value for $S_3(r)$. Note that $S_n(r)$ becomes more negative as the odd values
of $n$ increase.  Since the structure function of odd order is
determined from the difference of two terms of almost equal magnitude,
contributed from positive and negative values of $u_r$, a delicate balance
between the two terms is responsible for the behavior of the odd-order
structure functions.  The structure functions of even order are the
sum of those same terms, and do not rely on that intricate balance.
Thus reliable evaluation of the odd-ordered structure functions is
more difficult than evaluation of the even-ordered structure functions, and a smooth
change in the exponents of $S_n(r)$ with respect to varying $r$ in $n$,
is expected only for flows of extremely large Reynolds number. A way of
overcoming this difficulty for moderate Reynolds numbers is to employ the
generalized structure functions $\av{|u_r|^n}$ instead of
$S_n(r)$~\cite{rf:camussi1,rf:camussi2,rf:dhruva,rf:benzi2,rf:stolovitzky,rf:cao1,rf:belin,rf:grossmann}.
We thus introduce
\be
     \St_n(r)=\av{|u_r|^n}.  \label{5-2}
\ee
$\St_n(r)$ will now reflect the
change of the entire form of PDF, with respect to $u_r$, so that the scaling
exponents of $\St_n(r)$ are now expected to vary smoothly with $n$. It is expected
that the exponents of $S_n(r)$ and $\St_n(r)$ for an odd number of $n$ will be
identical for large Reynolds numbers. In what follows, we drop the symbol
tilde on $S_n$ so that no confusion will result.

When we plot $S_n(r)$ against $r_*$, it is difficult to read the exponent
$\zeta_n$ defined in the relationship
\be
     S_n(r) \sim r^{\zeta_n}.  \label{5-2'}
\ee
An example of this is shown in Fig. 7, where curves of $S_6(r)$ for Run 1 (decaying
turbulence, 12 samples) and for Run 4 (forced turbulence, 126 samples) are
plotted against $r_*$. The magnitude of both these quantities changes in such a way
that both curves agree with each other in the dissipative region. It should be noted
that the slope of $S_6(r)$ for decaying turbulence (Run 1) is smaller than that of
the same quantity for forced turbulence (Run 4). The local slope
$\d \log S_6(r)/\d \log r$ never assumes a constant value over an extended range,
as shown in Fig. 8. Often, for moderate Reynolds number flows, the ESS
method~\cite{rf:benzi1,rf:benzi2} is used instead of the above direct method.
Following the work of many authors ~\cite{rf:camussi1,rf:camussi2,rf:dhruva,rf:benzi1,rf:benzi2,rf:sreen,rf:stolovitzky,rf:cao1,rf:belin,rf:grossmann},
we have plotted $S_n(r)=\av{|u_r|^n}$ against $S_3(r)=\av{|u_r|^3}$ instead
of $\av{u_r^3}$, for the reason stated previously in this section. Figure 9 shows a
plot of $S_6(r)$ against the $S_3(r)$ corresponding to Fig. 7; the power law
tendency is clearly seen in this comparison.

The exponents of $S_n(r)$ are obtained from the graph of the local exponent of
$\log S_n(r)$ against $\log S_3(r)$. Combining $S_n(r) \sim r^{\zeta_n}$
and $S_3(r) \sim r^{\zeta_3}$, we have
\be
   S_n(r) \sim S_3(r)^{\zeta_n/\zeta_3}. \label{5-3}
\ee
Hence, the slope of the plot of $\log S_n(r)$ against $\log S_3(r)$ is
$\zeta_n/\zeta_3=\zeta_n$, since $\zeta_3$ is assumed to be unity. 

\begin{figure}
\epsfile{file=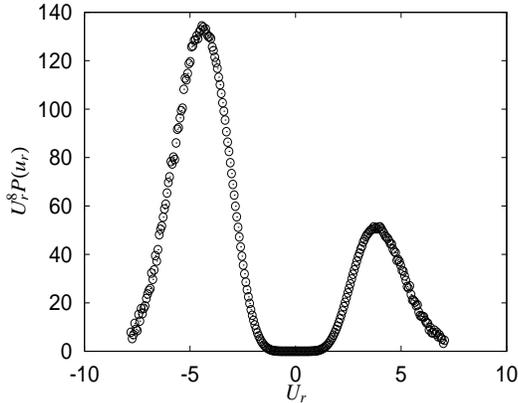,width=7.2cm,height=6.0cm}
\vspace{-8mm}
\caption{
The plot of $U_r^8 P(u_r)$ against $U_r$ at $r_*=20.8$ for Run 2, where
$U_r=u_r/\sqrt{S_2(r)}$.
}
\end{figure}
\begin{figure}
\epsfile{file=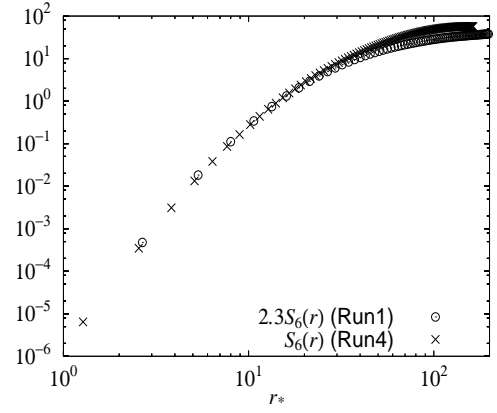,width=8.0cm,height=6.0cm}
\vspace{-8mm}
\caption{
The plot of $S_6(r)$ against $r_*$ for Run 1 (decaying turbulence) and
Run 4 (forced turbulence).
}
\end{figure}
\begin{figure}
\epsfile{file=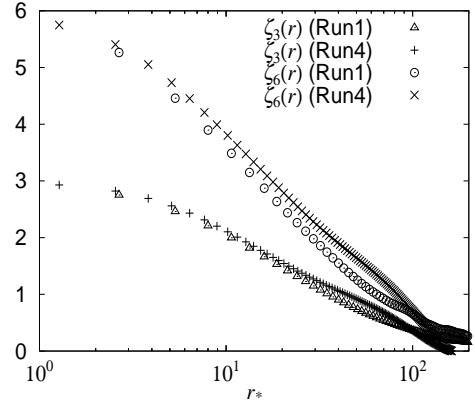,width=6.7cm,height=6.0cm}
\vspace{-8mm}
\caption{
The plot of the local slopes of $S_3(r)$ and $S_6(r)$ against $r_*$
for Run 1 (decaying turbulence) and Run 4 (forced turbulence). The power-law
dependence is not appreciable.
}
\end{figure}
\begin{figure}
\epsfile{file=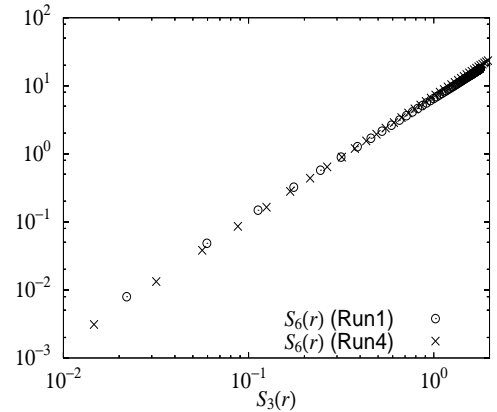,width=8.0cm,height=6.0cm}
\vspace{-8mm}
\caption{
The plot of $S_6(r)$ against $S_3(r)$ for Run 1 (decaying turbulence)
and Run 4 (forced turbulence). Here the power-law dependence is appreciable.
}
\end{figure}

\subsection{Decaying turbulence}

In this section, we will employ the local slope method, which has the advantage
that it requires less care in choosing where to draw the
appropriate straight fitting-line on the graph. In Fig. 10 the local slope of
$\log S_n(r)$ of Run 1 is depicted. Note that the abscissa in this plot is not
$\log S_3(r)$, but $\log r_*$, which is determined from $S_3(r)$. 
It is remarkable that all the local slopes take on constant values in the interval
$16.0 \leq r_* \leq 29.4$. This indicates that the scaling determined here is reliable.
Runs 2 and 3, using different Reynolds numbers and different realizations, have
almost the same exponents as Run 1, although the scaling-range differs slightly from
case to case. The local slopes $\zeta_n$ for Runs 1 to 3 are listed in Table
\ref{table:2} along with the scaling ranges. The last entry in the table shows the
standard numbers fitted by an SL model.~\cite{rf:sl} Observe that when the Reynolds
number increases, $\zeta_2$ increases while $\zeta_n, (n\ge 4)$ decreases.
The measured exponents in the decaying turbulence simulations are slightly
(but definitely) less than the standard numbers obtained in forced turbulence studies.

We will now turn to a discussion of the condition of isotropy and the extent of
non-stationarity in connection with the determination of the exponents of
structure function. The isotropic conditions for the second- and third-order
structure functions are guaranteed in the scaling range. We will now
seek a connection with the non-stationarity of the functions. Such effects are
inferred from the study of the KHK equation, or equivalently, from the
third-order structure function. The KHK relation of Run 1 (Fig. 4) shows that
the effect of the non-stationarity term $\partial D_{LL}(r)/\partial t$ is
not negligible in the scaling range $16.0 \leq r_* \leq 29.4$, which is the range in
which the scaling exponents of the structure functions for Run 1 are satisfactorily
obtained. Study of the results reveals that the non-stationary effect is $ 22.4 $\%
of the $(4/5)\ave r$ at $r_*=29.4$, which is the upper cutoff, while it is only
$ 8.2 $\% at $r_*=16.0$.
If the non-stationary effect is included in the total interaction, the energy
flux in the scale-space of $r$ is
\be
   \varepsilon(r) \equiv \ave \left( 1+\frac{15}{4 \ave r^5} \int^r_0
      \pa{D_{LL}(r',t)}{t}r'^4 \d r' \right),  \label{5-4}
\ee
which can be derived from eq. (\ref{4-2}). Since the integrand of the second term
on the right-hand side of the above equation is negative in decaying turbulence,
it follows that the function $\varepsilon(r)$ decreases as $r$ increases.

In order to see the effect of non-stationarity on higher-order structure
functions, it is useful to examine the equation for the $n+1$th moment of the increment
of the $i$th velocity component $w_i=u_i(\x_2)-u_i(\x_1)$, which is derived
in Appendix A as eq. (\ref{a-6'}). It is:
\ba
\paa{r_k} \av{w_k w_i^n} + n \av{w_i^{n-1} \paa{X_i} \delta p(\X,\r,t)} 
 \no\\
  & & \hspace{-55mm} =G_n^{(i)}(\r)-D_n^{(i)}(\r)+2 \nu \nabla_{\r}^2 \av{w_i^n}, \label{5-5}
\ea
where $G_n^{(i)}$ represents the non-stationarity effect
\be
  G_n^{(i)}(\r)=-\paa{t} \av{w_i^n},  \label{5-6}
\ee
and $D_n^{(i)}$ expresses the correlation of the dissipation rate and $w_i^n$,
which is written:
\be
       D_n^{(i)}(\r)=2n(n-1)\av{\varepsilon_i(\X,\r)w_i^{n-2}}.  \label{5-7}
\ee
Here $\varepsilon_i$ is the dissipation rate of the kinetic energy of the $i$th
velocity component at the two points $\X \pm \r/2$. It is written:
\ba
  \varepsilon_i(\X,\r)=\frac{\nu}{4} \left[
     |\nabla_{\X}w_i|^2+4|\nabla_{\r}w_i|^2
     \right] 
       \no \\
   & &\hspace{-55mm} =\frac{\nu}{2} \left[ |\nabla u_i(\X+\r/2)|^2+|\nabla
u_i(\X-\r/2)|^2
     \right]. \label{5-8}
\ea
A weighted energy flux can then be written as
\be
   \av{\varepsilon_i(\X,\r)w_i^{n-2}} \, [1-\rho_n^{(i)}(\r)], \label{5-9}
\ee
where
\be
   \rho_n^{(i)}(\r)= -\frac{1}{2n(n-1)}
   \frac{\paa{t} \av{w_i^n}}{\av{\varepsilon_i(\X,\r)w_i^{n-2}}}. \label{5-10}
\ee
Since $\varepsilon_i(\X,\r)$ is positive-definite, as can be established from its
definition (\ref{5-8}), and the PDF for $w_i$ is negatively skewed, then the
denominator in  $\rho_n^{(i)}(\r)$ is positive for even $n$ and negative for
odd $n$. On the other hand, the numerator is negative for even $n$, and
positive for odd $n$, because its amplitude decreases with time for decaying
turbulence. Thus $\rho_n^{(i)}(\r)$ is a positive quantity. The weighted flux is thus
smaller than the value without the inclusion of the non-stationarity effect.
Therefore it is possible that the non-stationarity effect yields a 
different scaling for the structure functions in question.

Although the KHK equation was not computed for Runs 2 and 3 (because the data
for the time derivative were not stored during measurement), we can estimate
the effect of non-stationarity by comparing the term 
$D_{LLL}-6\nu \partial D_{LL}/\partial r$ to $-(4/5) \ave r$, 
since the effect of decay is attributed to their difference. To
save on space in this paper, we do not include this figure in the text, but we
have noticed that the non-stationarity effect is not negligible in Runs 2 and 3
where larger Reynolds-number values are exhibited.

\begin{fulltable}[t]
\renewcommand{\arraystretch}{1.0}
\begin{center}
\begin{fulltabular}{@{\hspace{\tabcolsep}\extracolsep{\fill}}cccccccc} \hline
$ \enskip$    & Run 1 & Run 2 & Run 3 & Run 4 & Run 5 &  SL \\ \hline
condition & decay & decay & decay & forced & forced &   \\
{\scriptsize fitting region ${r_*}$ }& $16.0 \sim 29.4 $ & $11.7 \sim 26.0 $ &
$12.4 \sim
48.0$ & $11.5 \sim 28.1 $ & $10.2 \sim 30.6  $  & \\
$\zeta_2$  & 0.703 & 0.705   & 0.713  &  $0.690 $ & $0.692 $ & 0.696 \\
$\zeta_4$  & 1.266 & 1.264   & 1.252  &  $1.288 $ & $1.284 $ & 1.280 \\
$\zeta_5$  & 1.507 & 1.502   & 1.477  &  $1.555 $ & $1.546 $ & 1.548 \\
$\zeta_6$  & 1.724 & 1.718   & 1.680  &  $1.804 $ & $1.788 $ & 1.778 \\
$\zeta_7$  & 1.920 & 1.915   & 1.864  &  $2.037 $ & $2.011 $ & 2.001 \\
$\zeta_8$  & 2.096 & 2.095   & 2.035  &  $2.254 $ & $2.217 $ & 2.211 \\ \hline
\end{fulltabular}
\end{center}
\vskip -0.4cm
\caption{Scaling Exponents. The last entry, denoted as SL, shows the standard values
presented in ref. 3.}
\label{table:2}
\vskip -1.0cm
\end{fulltable}

\begin{figure}
\epsfile{file=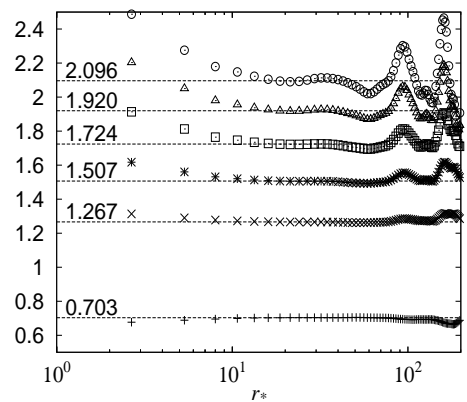,width=6.7cm,height=6.0cm}
\vspace{-8mm}
\caption{
The local slopes of the structure functions for Run 1; pluses stand for
$\zeta_2$, crosses for $\zeta_4$, stars for $\zeta_5$, squares for
$\zeta_6$, triangles for $\zeta_7$, and circles for $\zeta_8$.  The straight
dotted
lines are drawn with numbers averaged over the scaling region.
}
\end{figure}

\subsection{Forced turbulence}

We will now discuss the structure functions for Run 4 averaged over 126 samples
for approximately 50 eddy-turnover times.  The calculated local slopes have been
plotted against $r_*$ and are shown in Fig. 11.  The constancy of the slopes in
the interval $11.5 \leq r_* \leq 28.1$ is extremely impressive. Although the
Reynolds number is 70, the exponents are very close to the standard values.
Run 5, which has a flow of even higher Reynolds number, yields similar
results, as can be seen in Table \ref{table:2}.  Note that the scaling range
is rather narrow in Run 5, despite its larger Reynolds number, than in Run 4. We
suspect that the reason for this is that the sample number is not large enough
in Run 5, where we averaged over only 9 eddy turnover times, compared to the
50 eddy turnover times used in Run 4. In accord with current understanding
of the process, we see that longer runs yield wider scaling ranges.

We will now investigate the connection between isotropy and external forcing.
Note that for the scaling region used for this test, the isotropic conditions
are well satisfied, while the KHK equation is affected slightly by the
external-forcing terms. According to Fig. 5, the effect of external forces is
small in the above scaling range.  However, careful inspection of the numerical
results shows that the forcing term in KHK is 13.3\% of $(4/5) \ave r$ at $r_*=28.1$
and 2.3\% at $r_*=11.5$. This reveals that external forcing does not affect
the lower inertial region, which is where the constant local slopes begin,
but some effect is evident in the upper inertial region.

In forced turbulence, the equation for the $n+1$th moment of the increment
takes the form:
\ba
\paa{r_k} \av{w_k w_i^n} + n \av{w_i^{n-1} \paa{X_i} \delta p(\X,\r,t)}
     \no \\
  & & \hspace{-50mm} =H_n^{(i)}(\r)-D_n^{(i)}(\r)+2 \nu \nabla_{\r}^2 \av{w_i^n}, \label{5-11'}
\ea
where $H_n^{(i)}$ represents the contribution of the external force, and is written as
\be
  H_n^{(i)}(\r)=\frac{1}{9} n(n-1) \varepsilon_{in}(k_e r)^2 \av{w_i^{n-2}}.
\label{5-11}
\ee
(See Appendix A for a more complete explanation of this term.) The importance of
the external forces can then be estimated through the ratio of $H_n^{(i)}(\r)$ to
the transfer term on the left-hand side of eq. (\ref{5-11'}):\bd
  \frac{H_n^{(i)}(\r)}{\paa{r_k} \av{w_k w_i^n}} \sim (k_e r)^2 r^{\alpha_n},
    \label{5-12}
\ed
where $\alpha_n=1+\zeta_{n-2}-\zeta_{n+1}$, is an increasing function of $n$ and
approaches the constant value $2/3$, according to the SL model. This estimate
predicts that the influence of the external forces will diminish as the
order of the structure function increases. Consequently, the effect of the
external forces will not be significant in the higher-order structure functions.

\begin{figure}
\epsfile{file=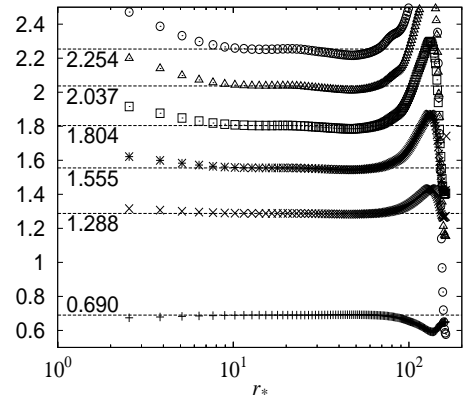,width=6.7cm,height=6.0cm}
\vspace{-8mm}
\caption{
The local slopes of the structure functions for Run 4. The plus symbols
represent $\zeta_2$, the crosses $\zeta_4$, the stars $\zeta_5$, the squares
$\zeta_6$, the triangles $\zeta_7$, and the circles $\zeta_8$.  The straight
dotted lines are drawn with numbers averaged over the scaling region.
}
\end{figure}

\subsection{Comparison}

The results of our careful simulations show the scaling exponents to be
definitely smaller in the decaying turbulence simulations than in the forced
turbulence simulations. Recently, Boratav and Pelz~\cite{rf:boratav} obtained the
scaling indices for simulation of decaying turbulence with $R_\lambda=82$, which agrees with the standard values. However, they did not employ
a method of computing the local slope of the structure functions
and identifying the scaling region with the constant local slope. They simply drew a best-fit
line on their graphed curve to estimate its slope. Such a method will predict a
different set of scaling indices depending on which region is chosen.

Let us now consider the reasons why the exponents are smaller for decaying
turbulence than for forced turbulence. One conceivable reason for this phenomenon
is that it is caused by the effect of the system's non-stationarity, as discussed
in \S 5.1.  One may argue that large Reynolds number flows do not suffer from
non-stationarity effects; however, since the magnitude of the non-stationarity
effects decreases only by a certain power law as $r$ decreases, the entire
inertial region should be affected by the influence of non-stationarity.

Another possible reason for this discrepancy is that the dissipative
structure may be different in decaying turbulence than in forced turbulence.  
Quantitatively,
\be
         \av{\varepsilon_i(\X,\r)w_i^{n-2}} \label{5-13}
\ee
may scale differently in decaying- and forced-turbulence environments.  
As is shown in Appendix B, the pressure gradient term in eq. (\ref{5-5}) is not affected 
by fluctuations at a far distance, and it obeys the same scaling relationship as the
inertial term. Hence the left-hand side of eq. (\ref{5-5}) is represented 
primarily by the inertial term. The essential equation for the inertial range is
\be
  (1+I^{(i)}_{n-1})\paa{r_k} \av{w_k w_i^n}
           =-2n(n-1)\av{\varepsilon_i(\X,\r)w_i^{n-2}},   \label{5-14}
\ee
where $I^{(i)}_{n-1}$ represents the contribution from the pressure gradient (see Appendix B). 
If the fluctuation of the dissipation rate is neglected, then the above equation
can be replaced by:
\be
(1+I^{(i)}_{n-1})\paa{r_k} \av{w_k w_i^n}=-\frac{2}{3} n(n-1) \ave \av{w_i^{n-2}},
\label{5-15}
\ee
which governs the K41 scaling. In real turbulence however, the dissipation
rate fluctuates in such a way that $[w_i(\X,r)]^{n-2}$ and $\varepsilon_i(\X,\r)$
are positively correlated. Under these circumstances we have intermittent
scaling, where the scaling exponents are smaller than their K41 values.

It is possible that the correlation of $[w_i(\X,r)]^{n-2}$ and $\varepsilon_i(\X,\r)$
differs for decaying turbulence and forced turbulence. This may happen 
because the dissipative structure differs for the two turbulences and the
non-stationarity effect may also affect the correlation (\ref{5-13}). Precise
analysis of the smaller exponents in decaying turbulence is a topic left for future study.

\section{Finite Reynolds Number Effect on the Scaling Exponents}

So far, we have computed the structure functions for decaying and forced
turbulence using a statistically sufficient number of samples, and have 
estimated the scaling exponents using the ESS method. In this section,
we discuss the behavior of the structure function. If one looks at
the plot of the structure function against $\log r_*$ (as for $\log S_6(r)$
in Fig. 7, for example,) it can be seen that the curve's shape may be approximated
in the scaling region by a quadratic form of $\log r_*$:
\be
   \log S_n(r)=A_n+B_n \log r_* -\frac{C_n}{2} (\log r_*)^2.  \label{6-1}
\ee
Here $A_n, \, B_n$ and $C_n$ are positive.  Note that as the Reynolds number, $R_{\lambda}$,
increases, the linear range of the above equation increases, indicating that $C_n$
decreases with $R_{\lambda}$. From the results of our studies, we suspect that $B_n$
varies little with changes in $R_{\lambda}$. The local exponents can then be computed
from
\be
   \zeta_n(r)=B_n -C_n \log r_*. \label{6-2}
\ee

As $r_*$ increases, the structure functions saturate (that is, become independent
of $r$), as can be seen from the plot of $S_6(r)$ in Fig. 7. This implies that 
$\zeta_n(r)$ goes to zero at the saturation scale. We then require that $\zeta_n(r)=0$
at $r=L_0$. Note that we do not necessarily think that $L_0$ is the same as the
external length. Consequently,
\be
     \log \frac{L_0}{\eta}=\frac{B_n}{C_n}=\mbox{const} \equiv M,  \label{6-3}
\ee
and substituting eq. (\ref{6-3}) into eq. (\ref{6-2}) yields
\be
   \zeta_n(r)=B_n \left( 1 - \frac{\log r_*}{M} \right). \label{6-4}
\ee
Since $C_n$ decreases with $R_{\lambda}$, $M$ will increase with $R_{\lambda}$.
Employing the ESS expression $\zeta_n=\zeta_n(r)/\zeta_3(r)$, eq. (\ref{6-4})
leads us to
\be
    \zeta_n=\frac{B_n(1 - \log r_*/M)}{B_3(1 - \log r_*/M)}=\frac{B_n}{B_3},
\label{6-5}
\ee
which is independent of $r_*$, so long as we stay in the scaling range.  This
explains why the ESS works so well for our simulations. In the limit of infinite
$R_{\lambda}$, $B_3\to 1$ and therefore, $B_n \to \zeta_n$.

We will now discuss the evidence supporting eq. (\ref{6-4}), relying on our
present simulations and other experiments employing large Reynolds numbers.
Figure 12 shows a plot of the local slopes $\zeta_n(r)$ against $r_*$ for Run 4, 
in which various straight lines fitted in the inertial region (where the scaling
exponents are to be determined) are extrapolated outside of the scaling region.
All the lines converge to a single point, $(10^M, \zeta_n=0)$, although
the data points for large $r_*$ values are suppressed to emphasize the
convergence. The relation (\ref{6-4}) is remarkably well satisfied with
$M=\log 143=2.16$. Other runs with forced turbulence show the same tendency
with $M=\log 188 =2.27$ for Run 5. It is clear from these results that $M$
increases as $R_{\lambda}$ increases. For the decaying turbulence of Run 1
using the same $R_{\lambda}$ as Run 4, $M=\log 112=2.05$ which is slightly
smaller than the corresponding value for Run 4. This is interpreted as likely
being due to a possibly larger $C_n$ value for decaying turbulence resulting
from the non-stationarity effect.

The linearity of $\zeta_n(r)$ with respect to $\log r_*$ is demonstrated in 
experiments for turbulent flows with Reynolds
numbers larger than the present study.~\cite{rf:sreen,rf:belin} It should be noted
though that the scale ranges where the linearity holds in these experiments
are located at larger scale values than in our studies.  Note that the local
slopes are shown for more restricted values of $n$ in those other experiments; for
example: $M=\log 2 \times 10^5=5.28$ in the flow of $R_{\lambda}=2000$,~\cite{rf:belin}
where $M$ is read from the plot of $\zeta_6(r)$ (see Fig. 2(d) in ref. 20). According
to the flow of $R_{\lambda}=19500$,~\cite{rf:sreen} $M \sim 11$ is estimated from
$\zeta_2(r)$ (from Fig. 4(a) in ref. 16), and $M \sim 10$ from $\zeta_6(r)$ (from
Fig. 4(b) in the same publication).  In order to see how $M$ varies with $R_{\lambda}$,
we create a log-log plot of $M$ against $R_{\lambda}$, as in Fig. 13.
From this we can establish the rough relationship: $M \sim R_{\lambda}^{0.3}$. We expect
this relationship to be confirmed in future work.

\begin{figure}
\epsfile{file=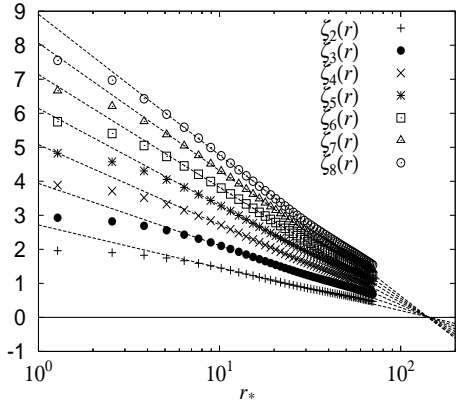,width=6.7cm,height=6.0cm}
\vspace{-8mm}
\caption{
The verification of relation (\ref{6-4}) for Run 4. The data
points for large values of $r_*$ are suppressed to emphasize that the
scaling lines converge to a single point at $(r_*=143, \zeta_n=0)$.
}
\end{figure}
\begin{figure}
\epsfile{file=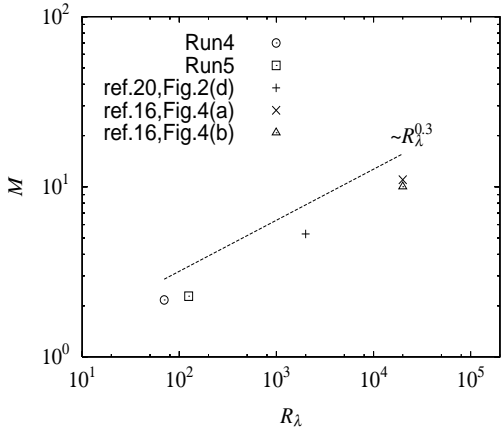,width=8.0cm,height=6.0cm}
\vspace{-8mm}
\caption{
The curve of $M$ against $R_{\lambda}$. A circle symbol represents the
data for Run 4, and a square the data for Run 5. A plus indicates the data for
the experiment of
$R_{\lambda}=2000$,~\cite{rf:belin} and a cross (derived from $\zeta_2(r)$) and a
triangle (derived from $\zeta_6(r)$) show the results for experiments of
$R_{\lambda}=19500$~\cite{rf:sreen}.  The broken line represents the
relation $M \sim R_{\lambda}^{0.3}$.
}
\end{figure}

\section{Summary and Discussion}

In this paper we have examined turbulence conditions 
to accurately determine the longitudinal structure function through 
the incompressible expressions of the second- and third-order structure 
functions and through the KHK relation. The examination of the isotropy for
the second- and third-order structure functions, in terms of the incompressibility
conditions, showed that the isotropy of these functions is well satisfied
when averages are taken over the ensemble of the initial field or over a sufficient
period. It was found that the non-stationarity effect due to the decaying
and forcing effects penetrates into the upper inertial range,
and the ESS method must be applied to measure the scaling exponents of the
structure functions on scales smaller than those of penetration.

The degree of anisotropy of turbulent flow will increase with the order of the
structure function, as can be inferred from our observation that the third-order
structure function is more anisotropic than the second-order function. The problem lies in
determining to what extent the anisotropy affects the higher-order 
structure functions.  It is difficult to numerically predict the anisotropic
effects, because the simple expressions for the isotropy of such systems, as related
in eqs. (\ref{2-1}) and 
(\ref{2-2}), have not yet been extended to include the higher-order structure functions.
Also, we do not yet know how the effect of non-stationarity
changes with the order of the structure functions.  This could possibly be estimated 
by examining numerically the factor defined in eq. (\ref{5-10}).

We have computed the scaling indices of the structure functions up to the eighth 
order, which we believe to be the maximum order for which accurate results are given
under the above considerations. It was found that the exponents are definitely smaller
for decaying turbulence than for forced turbulence. 
Further studies are necessary, however, to derive definite conclusions
either extending or disproving this observation for turbulence with higher-order
structure functions. Our limited current results for simulated decaying turbulence,
however, do indeed show the smaller numbers.

Since smaller exponents correspond to the more intermittent characteristics in
the turbulence, we see that decaying turbulence is more intermittent than forced
turbulence. It is widely recognized that a coherent structure with a very large
velocity gradient is of the form of a vortical tube which
has a diameter and length of the order of the Kolmogorov- and integral-lengths.
It is plausible that random forcing at a large scale acts to destroy the coherence
of such a structure over the integral scale. On the other hand, in decaying turbulence
there is no such mechanism to prevent the development of a coherent structure in
the velocity field, other than basic restrictions imposed by the computational
box-size or experimental geometry. For this reason, we observe that decaying turbulence
is more intermittent than the forced variety.

We have argued that the local-scaling exponents $\zeta_n(r)$ can be expressed as a
linear function of $\log r_*$, reflecting the finite nature of the Reynolds number.
When the Reynolds number becomes large, the correction term arising from the 
effect of the finite Reynolds number vanishes.

\section*{Acknowledgements}
The work of T.G. and T.N. presented here was supported by a Grant-in-Aid for
Scientific Research (C) 09640260 from The Ministry of Education, Science, Sports
and Culture of Japan, and by RIKEN.  We are grateful to T. Ochiai at NIT for
his assistance in the numerical computation involved in this work.

\appendix
\section{Derivation of Karman-Howarth-Kolmogorov Equation}

Although the derivation of the KHK equation is given in most standard books
dealing with turbulence,~\cite{rf:kh} we will derive this equation for a structure
function of arbitrary order and will include the effects of forcing. Note that the
KHK equation turns out to be nothing more than the equation for the second-order
structure function.

Let $\u_1$ and $\u_2$ be the velocity of the flow at locations $\x_1$ and $\x_2$.
The average velocity $\V$ and the relative velocity $\w$ are defined as
\bd
   \w(\x_2, \x_1)=\u_2-\u_1, \quad \V=(\u_2+\u_1)/2.
\ed
The equation for $\w$ is derived by taking the difference of the Navier-Stokes
equations at $\x_2$ and $\x_1$;
\ba
   & & \hspace{-7mm} \paa{t} \w(\x_2,\x_1)+(\u_2 \cdot \nabla_2+\u_1 \cdot \nabla_1) \w(\x_2,\x_1)
         \no\\
   & & \hspace{-5mm} = - \left( \paa{\x_2}+\paa{\x_1} \right) (p(\x_2)-p(\x_1))
         \no\\
   & & \hspace{-1.2mm} +\nu (\nabla_2^2+\nabla_1^2) \w(\x_2,\x_1) +\f(\x_2)-\f(\x_1), \label{a-1}
\ea
where $\f$ is an external random Gaussian force, $p$ is the pressure and $\nu$ is the
kinematical viscosity.  Notice that the variable $\x_2$ is held constant when the
partial derivative with respect to $\x_1$ is taken, and similarly for $\x_1$
when the partial derivative with respect to $\x_2$ is taken. Now, we introduce
the quantities
\bd
    \X=(\x_1+\x_2)/2, \quad \r=\x_2-\x_1.
\ed
Since
\bd
       \paa{\x_1}=\frac{1}{2}\paa{\X}-\paa{\r}, \quad
       \paa{\x_2}=\frac{1}{2}\paa{\X}+\paa{\r},
\ed
we are led to the expressions:
\ba
\u_2 \cdot \nabla_2+\u_1 \cdot \nabla_1 &=& \V(\X,\r,t) \cdot \nabla_{\X}+\w(\X,\r,t) \cdot \nabla_{\r},  \no \\
\nabla_2+\nabla_1 &=& \nabla_{\X}, \no \\
\nabla^2_2+\nabla^2_1 &=& \frac{1}{2} \nabla^2_{\X}+2\nabla^2_{\r}.  \no
\ea
Here $\w(\X,\r)$ is the velocity difference $\w(\x_1,\x_2)$ with a change of
variables from $\x_1, \, \x_2$ to $\X, \, \r$.

On substituting the above relations into eq. (\ref{a-1}),
the Navier-Stokes equation for the velocity difference becomes
\ba
   & & \left( \paa{t}+V_k(\X,\r,t) \paa{X_k} \right) w_i(\X,\r,t) 
      \no \\
   & & \hspace{2mm}  =-w_k(\X,\r,t) \paa{r_k} w_i(\X,\r,t) 
     -\paa{X_i} \delta p(\X,\r,t)\no \\ 
   & & \hspace{5.9mm}  +\nu \left(\frac{1}{2}\nabla_{\X}^2+2\nabla_{\r}^2 \right)w_i(\X,\r,t)         +\delta f_i(\X,\r,t), \no\\
          \label{a-2}
\ea
where
\ba
   & & \delta p(\X,\r,t)=p(\x_2,t)-p(\x_1,t), \no\\
   & & \delta \f(\X,\r,t)=\f(\x_2,t)-\f(\x_1,t). \no
\ea

Note that the following incompressible condition
\bd
\hspace{-15mm}  \paa{\X} \cdot \V=\frac{1}{2} \paa{\X} \cdot (\u(\x_2)+\u(\x_1))
\ed
\vspace{-3mm}
\bd
    =\frac{1}{2} \left( \paa{\x_2} \cdot \u(\x_2)+\paa{\x_1} \cdot \u(\x_1)
    \right)
    =0,
\ed
holds, because $\r$ is held constant when the partial derivative with respect
to $\X$ is taken.  Similarly,
\bd
  \paa{\X} \cdot \w=\paa{\r} \cdot \V=\paa{\r} \cdot \w=0.
\ed

Multiplying eq. (\ref{a-2}) by $w_i^{n-1}$, and reorganizing the dissipation term
slightly, we get
\ba
  & & \frac{1}{n} \paa t w_i^n+\frac{1}{n} \V(\X,\r,t) \cdot \nabla_{\X}
w_i^{n} \no \\
  & &  =-\frac{1}{n} \w \cdot \nabla_{\r} w_i^n - w_i^{n-1} \paa{X_i} \delta
p(\X,\r,t)
     \no\\
   & & \hspace{3.7mm} +\frac{\nu}{2} \nabla_{\X} \cdot (w_i^{n-1}
\nabla_{\X}w_i)+
        \frac{2}{n}\nu \nabla_{\r}^2 w_i^n \no \\
   & & \hspace{3.7mm} -2(n-1)\varepsilon_i(\X,\r)w_i^{n-2}+\delta f_i w_i^{n-1}. \label{a-3}
\ea
Here,
\ba
  \varepsilon_i(\X,\r)=\frac{\nu}{4} \left[
|\nabla_{\X}w_i|^2+4|\nabla_{\r}w_i|^2
     \right]\no\\ 
& & \hspace{-58mm} =\frac{\nu}{2} \left[ |\nabla u_i(\X+\r/2)|^2+|\nabla
u_i(\X-\r/2)|^2
     \right] \label{a-4}
\ea
is the summation of the energy-dissipation rate of the $i$th component of the
velocity field at the two points $\x_1$ and $\x_2$.

Let us now calculate the correlation of $\delta f_i$ and $w_i^{n-1}$, which appears
in eq. (\ref{a-3}),
\bd
  F_i(n)=\av{\delta f_i(t)w_i^{n-1}}_f,
\ed
where $\av{\cdot}_f$ signifies the average over Gaussian random forces.
Since the random force is simply delta correlated in time, 
and $\delta f_i(s) (s\le t)$ correlates with $\delta f_i(t)$ only at $t=s$, 
it is appropriate to use the approximation
\bd
   w_i(t)=\int_{-\infty}^t \delta f_i(s)\d s
\ed
in the computation of the correlation $\av{\delta f_i(t)w_i^{n-1}}_f$.
Substituting this result into one of the $n-1$ components of $w_i$, we have
\bd
   F_i(n)=(n-1) w_i^{n-2} \int_{-\infty}^t \av{\delta f_i(t)\delta f_i(s)}_f \d s.
\ed

In the present simulation, external forces are added randomly, once per time-step
$\Delta t$, so that
\bd
   F_i(n)=(n-1) w_i^{n-2}\av{\delta f^2_i}_f.
\ed

The correlation of random forces is calculated through
\ba
   & & \av{\delta f_i^2}_f=\sum_{\k,\q} \av{f_i(\k)f_i(\q)}_f
   \e^{i(\k+\q) \cdot \X} \no\\ 
& &  \hspace{17mm} \times \left( \e^{i \k \cdot \r/2}-\e^{-i \k \cdot \r/2}
\right)
   \left( \e^{i \q \cdot \r/2}-\e^{-i \q \cdot \r/2} \right) \no\\
   & & \hspace{11.5mm} =\sum_{\k} \av{f_i(\k)f_i^*(\k)}_f
     ( 1-\e^{-i \k \cdot \r})(1-\e^{i \k \cdot \r}) \no\\
   & & \hspace{11.5mm} =2\sum_{\k} \av{|f_i(\k)|^2}_f
     (1-\cos \k \cdot \r),  \no
\ea
where the homogeneity of the system is employed. Since the random forces are
limited to wavenumbers smaller than $1/r$, where $r$ is our scale of interest,
$\cos \k \cdot \r$ may be expanded in terms of $\k \cdot \r$.  
Furthermore, we can make use of the property that the external forces are
distributed in an isotropic way. Then
\bd
  \av{\delta f_i^2}_f
    =\sum_{\k} \av{|f_i(\k)|^2}_f (\k \cdot \r)^2
    =\frac{r^2}{3} \sum_{\k}\av{|f_i(\k)|^2}_fk^2.
\ed
When we introduce the definition
\bd
   k_e^2=\frac{\sum_{\k}\av{|f_i(\k)|^2}_fk^2}{\sum_{\k} \av{|f_i(\k)|^2}_f},
\ed
we have
\bd
  \av{\delta f_i^2}_f =\frac{1}{3} (k_e r)^2 \sum_{\k} \av{|f_i(\k)|^2}_f.
\ed
Since the energy-input rate $\varepsilon_{in}$ due to the external
forces is
\be
   \varepsilon_{in}=3\av{f_i^2}_f= 3\sum_{\k}\av{|f_i(\k)|^2}_f,
    \label{a-5}
\ee
we are led to the relation
\be
   F_i(n)=\frac{1}{9}(n-1)\varepsilon_{in}(k_e r)^2 w_i^{n-2}.  \label{a-6}
\ee
Substituting eq. (\ref{a-6}) into eq. (\ref{a-3}) and taking the ensemble 
average over the velocity field, we have
\ba
  & & \paa t \av{w_i^n} =-\paa{r_k} \av{w_k w_i^n} - n \av{w_i^{n-1} \paa{X_i}
    \delta p(\X,\r,t)} \no\\
   & & \hspace{16.4mm} + 2 \nu \nabla_{\r}^2 \av{w_i^n}-
2n(n-1)\av{\varepsilon_i(\X,\r)w_i^{n-2}} \no\\
& & \hspace{16.4mm} +\frac{1}{9}n(n-1)\varepsilon_{in}(k_e r)^2 \av{w_i^{n-2}}.
    \label{a-6'}
\ea
Note that we have not made the summation over $i$ at this point.

Let us now focus on the KHK relation, which is derived by substituting $n=2$ into
eq. (\ref{a-6'}) and making the summation over $i$. We then have:
\ba
    \paa t \sum_i \av{w_i^2} + \paa{r_k} \sum_i \av{w_k  w_i^2} \no\\
  & &\hspace{-45mm}   =2\nu \nabla_{\r}^2 \sum_i \av{w_i^2}-4 \sum_i \av{\varepsilon_i(\X,\r)}+\frac{2}{3}\varepsilon_{in}(k_e r)^2.  \no\\
     \label{a-7}
\ea
Making use of the relations from ~\cite{rf:kh}
\bd
   \sum_i \av{w_i^2}=\frac{1}{r^2} \paa{r} r^3 D_{LL},
\ed
\vspace{-3mm}
\bd
   \sum_i \av{w_k w_i^2}=\frac{r_k}{3r^4} \paa{r}r^4 D_{LLL},
\ed
\vspace{-3mm}
\bd
   \sum_i \av{\varepsilon_i(\X,\r)}=\ave,
\ed
we arrive at
\ba
  \frac{1}{r^2} \paa{r} r^3 \paa{t} D_{LL}+\frac{1}{3r^2} \paa{r} \left[
   \frac{1}{r} \paa{r} r^4 D_{LLL}\right] \no\\
& & \hspace{-50mm} =\frac{2\nu}{r^2} \paa{r} \left[
   \frac{1}{r} \paa{r} r^4 \paa{r}D_{LL}\right]-4\ave+\frac{2}{3}
     \varepsilon_{in}
    (k_e r)^2. \no
\ea
First, we multiply the above equation by $3r^2$, and then integrate it over
$r$ from 0 to $r$.  Second, we multiply the result by $r$, and integrate it
again over $r$.  We then get
\ba
\hspace{-10mm}
  D_{LLL}=-\frac{4}{5}\ave r + 6\nu \pa{D_{LL}}{r} \no\\
& & \hspace{-26.5mm} -\frac{3}{r^4}
    \int^r_{0}\pa{D_{LL}}{t}r'^4 \d r'+\frac{2}{35} \varepsilon_{in} (k_e r)^2 r,
   \label{a-9}
\ea
for $(k_e r)^2 \ll 1$. Note that the last term on the right-hand side of 
(\ref{a-9}), for arbitrary $r$, is given by:
\ba
   \frac{4 \varepsilon_{in} r}{\sum_{\k} \av{|f_i(\k)|^2}}
    \sum_{\k} \av{|f_i(\k)|^2} \no\\
& & \hspace{-35mm} \times\left( \frac{1}{5}+\frac{3}{k^3 r^3} \sin kr
   +\frac{9}{k^4 r^4} \cos kr-\frac{9}{k^5 r^5}\sin kr \right).\no\\  \label{a-10}
\ea

\section{Contribution of the Pressure Gradient Term}

The equation for the structure function of arbitrary-order has been derived above.
Each term appearing in the equation acts locally in space except for the
pressure-gradient term. The pressure is usually considered to be a long-ranged
influence, because it is related to the far velocity-fields through the Poisson kernel.
Consequently, the same is commonly thought to be true for pressure-gradient
forces.  Here we show that the pressure-gradient term is not as long-ranged as thought.
The fact that the pressure-gradient term has a local nature was mentioned previously by
L'vov and Procaccia~\cite{rf:lp}, but the argument can be better illustrated if we use
the results in the Appendix A.

Let us take the divergence of eq. (\ref{a-2}) with respect to $X_i$.  Since the
external forces are divergence-free, we are led to the equation:
\ba
   \nabla^2_{\X} \delta p(\X,\r,t) =-\paa{X_j} \biggl[
     w_k(\X,\r,t)\paa{r_k} w_j(\X,\r,t) \no \\
  & & \hspace{-60mm} +V_k(\X,\r,t)\paa{X_k}w_j(\X,\r,t)\biggr].
     \label{b-1}
\ea
The source term can be simplified as follows.  The incompressibility condition
reduces the source term to
\bd
    -\pa{w_k(\X,\r,t)}{X_j}\pa{w_j(\X,\r,t)}{r_k}
    -\pa{V_j(\X,\r,t)}{X_k}\pa{w_k(\X,\r,t)}{X_j},
\ed
where the interchanging of $(j \leftrightarrow k)$ has been done in the second term.
Since
\ba
   \paa{X_k} V_j(\X,\r,t) &=& \frac{1}{2} \paa{X_k} \left[
u_j(\X+\r/2)+u_j(\X-\r/2)
    \right] \no\\
    &=& \paa{r_k} \left[ u_j(\X+\r/2)-u_j(\X-\r/2) \right]\no\\
    &=& \paa{r_k}
w_j(\X,\r,t), \no
\ea
the source term becomes
\ba
\hspace{-5mm}
   -2\pa{w_k(\X,\r,t)}{X_j}\pa{w_j(\X,\r,t)}{r_k} \no \\
& &\hspace{-25mm}  =-2 \paa{X_j} \paa{r_k} w_k(\X,\r,t)w_j(\X,\r,t). \no
\ea
The pressure-gradient term then reduces to:
\ba
\hspace{-7mm}
  \paa{X_i} \delta p(\X,\r)=\int \d \X' K_{ij}(\X,\X') \no\\
   & & \hspace{-30mm}\times \paa{r_k}
     w_k(\X',\r,t)w_j(\X',\r,t),
\ea
where
\be
   K_{ij}(\X,\X')=\frac{1}{2\pi} \paa{X_i}\paa{X_j} \frac{1}{|\X-\X'|}.
\ee
The pressure-gradient term appearing in eq. (\ref{a-6'}) becomes
\ba
   E_{n-1}(w_i(r)) &\equiv & n\av{w_i^{n-1} \paa{X_i} \delta p} \no\\
     &=& n\int \d \R K_{i\alpha}(\R)B^{(n-1)}_{\alpha}(\R,\r,t),
\label{b-4}\\
    B^{(n-1)}_{\alpha}(\R,\r,t) &=&\biggl\langle w_i^{n-1}(\X,\r,t)
            \paa{r_{\beta}}  w_{\beta}(\X+\R,\r,t) \no \\
& & \times w_{\alpha}(\X+\R,\r,t)\biggr\rangle, \label{b-4'}
\ea
where the summation is taken only over the repeated Greek variables, and
$K_{ij}(\R)$ is a dipole-type interaction:
\be
   K_{ij}(\R)=\frac{1}{2\pi R^3} \left[ 3 \frac{R_i R_j}{R^2} -\delta_{ij}
\right].
   \label{b-5}
\ee
The average value (\ref{b-4'}) does not contain the following contribution,
\bd
   \av{w_i^{n-1}(\X,\r,t)} \av{ \paa{r_{\beta}} w_{\beta}(\X+\R,\r,t)
          w_{\alpha}(\X+\R,\r,t)},
\ed
since the latter average vanishes due to the fact that the divergence of
the second-order structure function is zero. Therefore, the two quantities
at $\X$ and $\X+\R$, that is:
\bd
   w_i^{n-1}(\X,\r,t) \quad \mbox{and} \quad \paa{r_{\beta}}
                         w_{\beta}(\X+\R,\r,t)w_{\alpha}(\X+\R,\r,t)
\ed
must be correlated.  It is certain that such a correlation function must
decrease with increasing separation $\R$ for $R \geq r$.

For $|\r|$ in the inertial range, the integral of eq. (\ref{b-4}) may be
written as
\be
  \left(\int_{R < r} \d\R+\int_{R > r} \d\R\right)
                         K_{i\alpha}(\R)B^{(n-1)}_{\alpha}(\R,\r). \label{b-4a}
\ee
The first integral, $B^{(n-1)}_{\alpha}(\R,\r)$, can be Taylor-expanded in
$|R/r|$ as
\be
  B^{(n-1)}_{\alpha}(\R,\r)=B^{(n-1)}_{\alpha}(0,\r)
          +\nabla_{\R}B^{(n-1)}_{\alpha}(0,\r)\cdot\R+\cdots. \label{b-4b}
\ee
When this form is substituted into the first integral of eq. (\ref{b-4a}),
the first term of eq. (\ref{b-4b}) vanishes due to the symmetry of the system:
\bd
   \int \d\R \frac{1}{2\pi R^3} \left[ 3 \frac{R_i R_j}{R^2}-\delta_{ij} \right]
    =0.
\ed
The other terms will make their contributions mostly around $R \sim r$.

Examining the second integral in eq. (\ref{b-4a}), we may
reasonably assume that the correlation $B^{(n-1)}_{\alpha}(\R,\r)$ behaves as
$(R/r)^{-\delta}$, for $R/r>1$ with $\delta>0$. Thus the contribution from the upper
boundary of the second integral of eq. (\ref{b-4a}) can be neglected, because
\be
\int \d\R \frac{1}{2\pi R^3} \left[ 3 \frac{R_i R_j}{R^2}
     -\delta_{ij} \right] R^{-\delta}\to 0 \qquad \mbox{ as } R\to \infty.
\ee

Since this integral is dominated by scales of $R$ comparable to $r$,
$B^{(n-1)}_{\alpha}(\R=\r,\r)$ can be approximated as a homogeneous 
function of $r$,
\bd 
B^{(n-1)}_{\alpha}(r\nnu,\r)
      \approx r^{\phi}L^{(n-1)}_{\alpha}(\nnu,\mmu), \label{b-4c}
\ed
where $\nnu=\R/R, \mmu=\r/r$ and $L^{(n-1)}_{\alpha}(\nnu,\mmu)$ is a geometric 
function of $\nnu$ and $\mmu$. 
$\phi$ is the same as the scaling exponent of $\av{w_i^{n-1}\paa{r_{\beta}}w_{\beta} w_\alpha}$.

Now $E_{n-1}(w_i(r))$ can be evaluated as  
\ba
E_{n-1}(w_i(r))&\approx& r^{\phi}nJ_i^{(n-1)}(\mmu),             \label{b-4d}\\
\hspace{-20mm} J_i^{(n-1)}(\mmu)&=&\frac{1}{2\pi}\int \d\Omega(\nnu)
                 \left(3\nu_i\nu_{\alpha}-\delta_{i\alpha}\right)  
                             L^{(n-1)}_{\alpha}(\nnu,\mmu).     \no\\
   \label{b-4e}
\ea
This development implies that the greatest contribution to the integrals
(\ref{b-4}) comes from the range of $R$ comparable to $r$, meaning
that the dominant contribution to the pressure-gradient term is from the near
correlation on the order of $R \sim r$. As far as the scaling of the 
pressure-gradient term in $r$ is concerned, the contribution from the pressure-gradient
in eq. (\ref{a-6'}) obeys the same scaling as the inertial term
$\paa{r_{\alpha}} \av{w_{\alpha}w_i^n}$. We may then write: 
\be
    E_{n-1}(w_i(r))=I^{(i)}_{n-1}(\mmu)\paa{r_{\alpha}}\av{w_{\alpha} w_i^n},         \label{b-4f}
\ee
where $I^{(i)}_{n-1}$ is a function of the order one, depending on $\mmu$.
The factor $n$ on the right hand side of eq. (\ref{b-4d}) is absorbed into the exponent of $\paa{r_{\alpha}}\av{w_{\alpha} w_i^n}$.


\begin{thebibliography}{99}
\bibitem{rf:my} A.S. Monin and A.M. Yaglom:
{\it Statistical Fluid Mechanics}, volume II  (MIT Press, Cambridge, Mass.,
1973),
W.D. McComb:
{\it The physics of fluid turbulence}  (Oxford University Press, New York,
1990),
U. Frisch: {\it Turbulence} (Cambridge University Press, Cambridge, England,
1995). 
\bibitem{rf:k} A.N. Kolmogorov:
Dokl. Akad. Nauk. SSSR {\bf 30} (1941) 299. 
\bibitem{rf:sl} In order to see numerically how the exponents deviate from K41,
refer to Z.S. She and E. Leveque:
Phys. Rev. Lett. {\bf 72} (1994) 336. 
\bibitem{rf:camussi1} R. Camussi, D. Barbagallo, R. Guj and F. Stella:
Phys. Fluids {\bf 8} (1996) 1181. 
\bibitem{rf:camussi2} R. Camussi and R. Benzi:
Phys. Fluids {\bf 9} (1997) 257. 
\bibitem{rf:dhruva} B. Dhruva, Y. Tsuji and K.R. Sreenivasan:
Phys. Rev. E\,{\bf 56} (1997) R4928. 
\bibitem{rf:noullez} A. Noullez, G. Wallace, W. Lempert, R.B. Miles and U. Frisch:
J. Fluid Mech. {\bf 339} (1997) 287. 
\bibitem{rf:boratav} O.N. Boratav and R.B. Pelz:
Phys. Fluids {\bf 9} (1997) 1400. 
\bibitem{rf:chen} S. Chen, K.R. Sreenivasan, M. Nelkin and N. Cao:
Phys. Rev. Lett. {\bf 79} (1997) 2253. 
\bibitem{rf:mtw} F. Moisy, P. Tabeling and H. Willaime:
Phys. Rev. Lett.  {\bf 82} (1999) 3994. 
\bibitem{rf:kh} A detailed presentation of this equation can be found in Monin
and Yaglom in ref. 1, and T. Gotoh: {\it Fundamental Turbulence Theory}
(Asakura, Tokyo, 1998) 1st ed., [in Japanese]. 
\bibitem{rf:benzi1} R. Benzi, S. Ciliberto, R. Tripiccione, C. Baudet, F. Massailoli
and S. Succi: Phys. Rev. E {\bf 48} (1993) R29. 
\bibitem{rf:benzi2} R. Benzi, S. Ciliberto, C. Baudet and G.R. Chavarria:
Physica D\,{\bf 80} (1995) 385. 
\bibitem{rf:lp} V.S. L'vov and I. Procaccia:
Phys. Rev. Lett. {\bf 77} (1996) 3541. 
\bibitem{rf:k2} A. N. Kolmogorov:
Dokl. Akad. Nauk. SSSR {\bf 32} (1941) 1. 
\bibitem{rf:sreen} K.R. Sreenivasan and B. Dhruva:
Prog. Theor. Phys. Suppl. {\bf 130} (1998) 103. 
\bibitem{rf:danaila} L. Danaila, F. Anselmet, T. Zhou and R.A. Antonia:
J. Fluid Mech. {\bf 391} (1999) 359. 
\bibitem{rf:stolovitzky} G. Stolovitzky and K.R. Sreenivasan:
Phys. Rev. E\,{\bf 48} (1993) R33. 
\bibitem{rf:cao1} N. Cao, S. Chen and Z.S. She:
Phys. Rev. Lett. {\bf 76} (1996) 3711. 
\bibitem{rf:belin} F. Belin, R. Tabling and H. William:
Physica D\,{\bf 93} (1996) 52. 
\bibitem{rf:grossmann} S. Grossmann, D. Lohse and A. Reeh:
Phys. Rev. E\,{\bf 56} (1997) 5473. 


\end{thebibliography}
\end{document}